\documentclass[preprint,aps,prb,showpacs,preprintnumbers,amsmath,amssymb,groupedaddress]{revtex4}
\usepackage{graphicx}
\begin{document}

\draft
\title {Electric Field Control of Shallow Donor Impurities in Silicon}
\author{A. S. Martins, R. B. Capaz, and Belita Koiller}
\affiliation{Instituto de F\'{\i}sica, Universidade Federal do Rio de 
Janeiro, 
Cx.P. 68.528, 21945-970, RJ, Brazil}
\date{\today}

\begin{abstract}
We present a tight-binding study of donor impurities in Si, 
demonstrating the adequacy of this approach for this problem
by comparison with effective mass theory and experimental results.
We consider the response of the system to an applied electric field:
donors near a barrier material and in the presence of an uniform 
electric field may undergo two different ionization regimes according 
to the distance of the impurity to the Si/barrier interface. We show 
that for impurities $\sim 5$ nm below the barrier, adiabatic ionization 
is possible within switching times of the order of one picosecond, 
while for impurities $\sim 10$ nm or more below the barrier, no 
adiabatic ionization may be carried out by an external uniform electric 
field. Our results are discussed in connection with proposed Si:P 
quantum computer architectures.
\end{abstract}

\maketitle

\section{Introduction}

Simple donors in Si have recently become the subject of renewed interest 
due to proposals of quantum computer architectures in which P donors in Si play 
the role of qubits.\cite{kane,vrijen,skinner} Logic operations in such 
architectures involve the response of the bound electron wavefunctions to 
voltages applied to a combination of metal gates separated by a barrier 
material (e.g. SiO$_2$) from the Si host. The so-called A-gate, placed above 
each donor site, pulls the electron wavefunction away from the donor, aiming at 
partial reduction \cite{kane} or total cancelation \cite{skinner} of the 
electron-nuclear contact coupling in architectures where the qubits are the 
$^{31}$P nuclear spins. In a related proposal based on the donor electron 
spins as qubits, \cite{vrijen} the gates drive the electron wavefunction into 
regions of different $g$-factors, allowing the exchange coupling between 
neighboring electrons to be tuned. Ideally, electric-field control over the 
donor electron wavefunction requires all operations to be performed in the 
adiabatic regime \cite{messiah}, which sets a lower bound for the time scales 
involved in such processes.

Recent studies have demonstrated that the tight-binding (TB) approach, 
traditionally adopted for deep levels,\cite{hjalmarson} provides a valid 
description for intermediate \cite{form1,form2} and shallow levels
\cite{martins02} in semiconductors.  Impurity states are calculated from a 
sequence of supercell sizes and a finite-size analysis which provides 
extrapolation to the bulk limit. Also, electric-field effects may be 
easily incorporated within the TB scheme \cite{graf}, allowing estimates of 
switching times in electric-field-tunable devices \cite{ribeiro}.  In
this work we present a TB description for donors in Si, aiming at a 
physical description of the relevant properties involved in the A-gate 
operations mentioned above.

Donors in Si have been extensively and successfully investigated within 
the effective mass theory (EMT), \cite{emt} thus providing a preliminary test 
for the TB approach by comparison of wavefunctions predicted by  
the two formalisms. This comparison is presented in the next section. In 
Sec.~\ref{sec:field} we explore a simplified model of the A-gate operations 
in the Kane quantum computer proposal \cite{kane} by considering the Si:P 
system under an uniform electric field and near a barrier. In Sec.~\ref
{sec:times} we discuss operation times and restrictions imposed by the donor 
positioning with respect to the Si/barrier interface in connection with the 
adiabaticity of the A-gate operations. Our summary and conclusions are 
presented in Sec.~\ref{sec:end}.

\section{TB description for donors in Silicon}
\label{sec:TB}
\subsection{Formalism}

The TB Hamiltonian for the impurity problem is written as \cite{form1}
\begin{equation} 
H=\sum\limits_{ij}\sum_{\mu 
\nu}h_{ij}^{\mu\nu}c_{i\mu}^{\dagger}c_{j\nu}+\sum_
{i,\nu} U(r_i)c_{i\nu}^{\dagger}c_{i\nu} \label{hamilton} 
\end{equation}
where $i$ and $j$ label the atomic sites, $\mu$ and $\nu$ denote the 
atomic orbitals and $r_{i}$ is the distance of the site $i$ to the impurity 
site. The matrix elements $h_{ij}^{\mu\nu}$ define all the on-site energies 
and first and second neighbors hoppings for the bulk material. The donor 
impurity potential $U(r_i)$ is described by a screened Coulomb potential 
($\epsilon=12.1$ for Si)
\begin{equation} U\left(r_{i}\right)=-\frac{e^{2}}{\epsilon r_{i}}. 
\label{potential} 
\end{equation} 
At the impurity site $\left(r_{i}=0\right)$, the perturbation potential is 
assigned the value $-U_{0}$, a parameter describing central cell effects 
characteristic of the substitutional species. In the present calculations, 
$U_{0}$ was kept as an adjustable parameter (previous estimates for this 
parameter\cite{form1} are of the order of one to a few eV). We adopt here 
the $sp^{3}s^{\ast}$ TB parametrization for Si proposed by Klimeck {\it et 
al},\cite{boysi} which includes first and second neighbors interactions. 
Inclusion of hopping matrix elements up to second neighbors provides a good
description of the effective masses at the conduction band minima. This 
parametrization gives the $k$-space positions of the six band minima at the 
six equivalent points along the $\Delta$ lines, at $\Delta_{min}=0.75 
(2\pi/{\rm a})$, where $\rm{a} = 5.431$\AA~is the conventional cubic lattice 
parameter for Si. We do not include spin-orbit corrections in our calculations.

The eigenstates of $H$ are determined for a system where a single impurity 
is placed in a cubic supercell containing $N=8L^{3}$ atoms arranged in the 
diamond structure, where $L$ is the length of the supercell edge in units of 
${\rm a}$. The supercells are subject to periodic boundary conditions, and 
full numerical diagonalization can be performed for $L\lesssim 6$. Much larger 
supercells\cite{martins02} (up to $10^6$ atoms) may be treated within a variational 
scheme\cite{araujo} where the ground state wavefunction and binding energy $E_{L}$ 
for a donor level is obtained by minimizing the expectation value of 
$\left\langle\Psi\left|\left(H-\varepsilon_{ref}\right)^{2}\right|\Psi 
\right\rangle$. For the donor ground state, $\varepsilon_{ref}$ is a reference 
energy chosen well within  the gap, but nearest to the conduction band minimum, 
and excited states are obtained by tuning $\varepsilon_{ref}$ towards the conduction 
band edge. Finite size scaling allows extrapolation to the bulk limit ($L\to\infty$) 
according to the {\it ansatz}~\cite{form1,form2} 
\begin{equation} 
E_{L}=E_{b} + \widetilde{E}e^{-L/\lambda}, 
\label{ansatz}
\end{equation} 
where $E_b$ is the binding energy for a single donor in the bulk.

The eigenfunctions of (\ref{hamilton}) written in the basis of atomic 
orbitals $|\phi_{\nu}(\mathbf{r}-\mathbf{R}_i)\rangle$ are given by 
$|\Psi_{TB} (\mathbf{r})\rangle = \sum_{i\nu} a_{i\nu}|\phi_{\nu}
(\mathbf{r}-\mathbf{R}_i)\rangle $ where the expansion 
coefficients $a_{i\nu}$ give the probability amplitude of finding the 
electron in the orbital $\nu$ localized at $\mathbf{R}_i$. We do not include 
explicit expressions for the atomic orbitals; the overall charge distribution 
is conveniently described through the TB envelope function squared,\cite{tania} 
\begin{equation}
|\Psi_{EF}({\mathbf{R}_i})|^2 = \sum_{\nu} |a_{i\nu}|^2.
\label{envfunc} 
\end{equation}

\subsection{Donor ground state}

\begin{figure}
\includegraphics[height=5.5in]{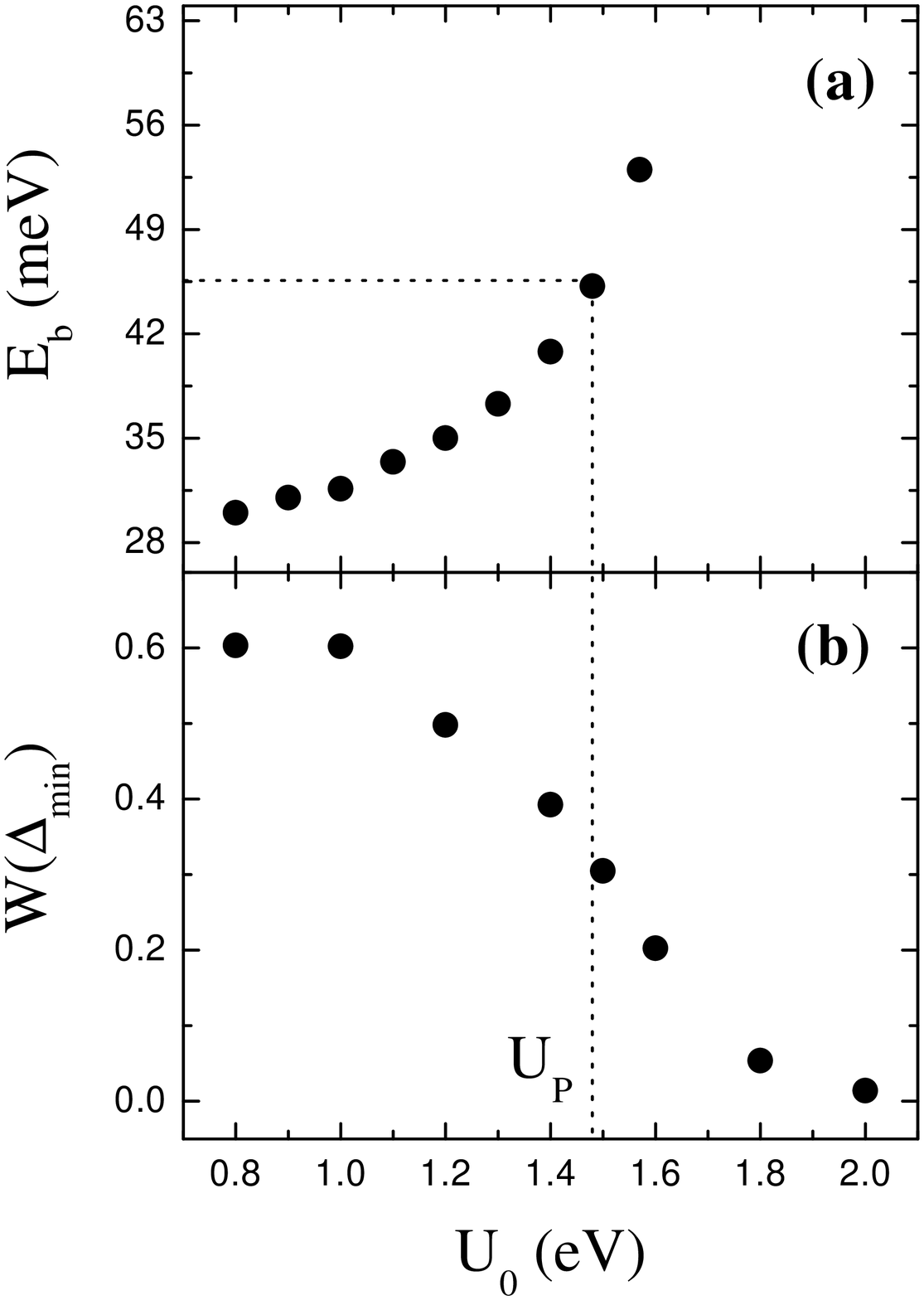}
\caption{\textbf{(a)} Binding energy of the ground impurity state as a function 
of the on-site perturbation strength $U_0$, obtained from the $L\to\infty$ 
extrapolation ansatz. The dotted line indicates the value $U_0 = U_P$ that 
reproduces the experimental Si:P $A_1$ state binding energy. \textbf{(b)} 
Calculated spectral weight at the conduction band edge for the ground state.
Notice that as the perturbation $U_0$ becomes weaker, $E_b$ approaches the
EMT binding energy, while $W$ does not approach 1.
\label{fig1}}
\end{figure}

In the proposed TB model, the only free parameter is related to the 
on-site value for the impurity potential, $U_0$. In Fig.~\ref{fig1}(a) we 
present the converged $(L\to \infty)$ binding energy of  the lowest donor state 
as a function of $U_0$. We also characterize the donor ground state by its 
orbital averaged spectral weight\cite{tania} at $\Delta_{min}$
\begin{equation}
W(\Delta_{min})=\frac{2}{N}\sum_{\mu=1}^6\sum\limits_{ij\nu}e^{i{\bf{k}_\mu}
\cdot({\bf R}_{i}-{\bf R}_{j})} a_{i\nu}a_{j\nu},
\label{espectral}
\end{equation}
where $N$ is the number of atomic sites in the supercell, and the first summation 
is over the six equivalent $\mathbf{k_\mu}$ at the conduction band minima. This 
quantity is plotted in Fig.~\ref{fig1}(b) as a function of $U_0$.
 
We determine the value of $U_0$ so that the binding energy of the donor 
results to be in good agreement with the experimental value which, 
for P in Si, is $E_b=45.6$~meV. As indicated in Fig.~\ref{fig1}, $U_{0}=U_{\rm P}
\cong 1.48$ eV gives the correct binding energy for the P donors in Si. 
This value for $U_0$ is used in the calculations below. 

\subsection{Comparison with EMT}

\begin{figure}
\includegraphics[height=6in]{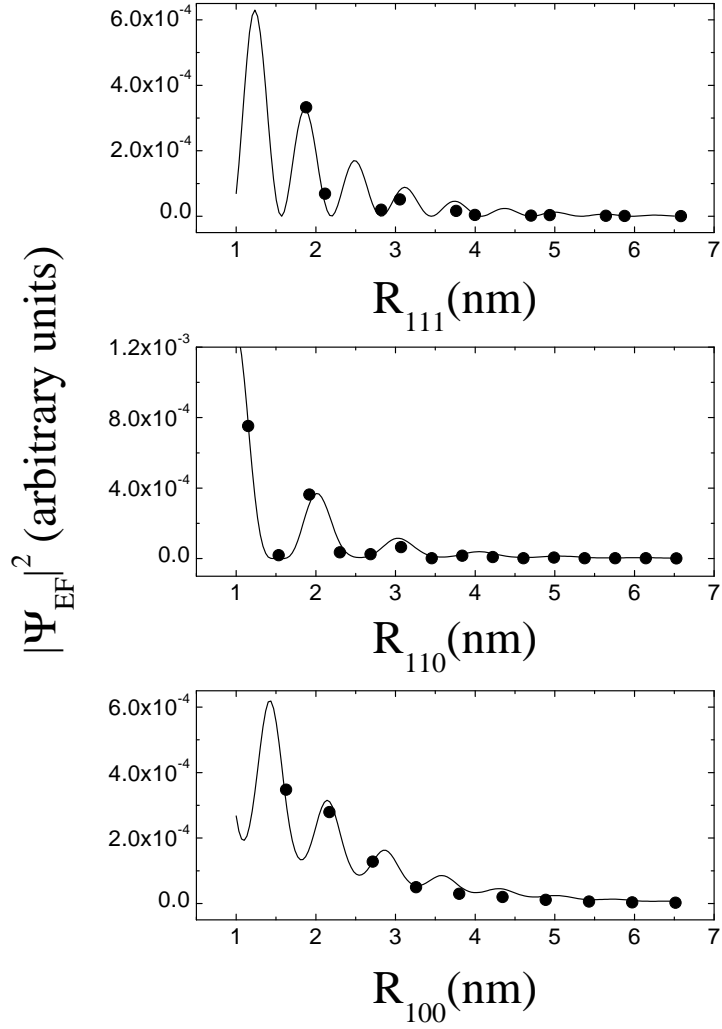}
\caption{The dots give the TB envelope function squared for the lowest impurity 
state along three high-symmetry directions. The lines are the corresponding 
effective mass $|\psi|^2$ results. Note that the TB approach captures the 
oscillations of the EMT wave function in the asymptotic region.
\label{fig2}}
\end{figure}

EMT exploits the duality between real and reciprocal space, where 
delocalization in real space leads to localization in $k$-space, e.g. for 
shallow donors around the $k$-vector at the minimum of the conduction 
band. Within EMT in its simplest formulation,\cite{emt} the ground state 
for donors in Si is six-fold degenerate, due to the six-fold 
degeneracy of the Si conduction band. Valley-orbit interactions \cite
{baldereschi} lead to a non-degenerate ground state wavefunction of $A_1$ 
symmetry,\cite{foot-symmetry}
\begin{equation}
\psi ({\bf r}) = \frac{1}{\sqrt{6}}\sum_{\mu = 1}^6 F_{\mu} ({\bf r})
\phi_\mu(\bf r)\,,
\label{eq:sim}
\end{equation}
where $\phi_\mu({\bf r}) = u_\mu({\bf r})e^{i {\bf k}_{\mu}\cdot{\bf r}}$ 
are the pertinent Bloch wave functions, and the envelope functions given by 
(e.g. for $\mu = z$)\cite{emtsi}  
\begin{equation}
F_{z} ({\bf r}) = \frac{1}{\sqrt{\pi a^2 b}}\ e^{-[(x^2+y^2)/a^2 + 
z^2/b^2]^
{1/2}} \,.
\label{eq:envelope}
\end{equation}
The effective Bohr radii for Si from a variational calculation are $a=2.51$~nm 
and $b=1.44 $~nm.\cite{koiller01} In Fig.~\ref{fig2} we present the TB envelope 
function squared calculated from (\ref{envfunc}) along three symmetry 
directions with the corresponding EMT results obtained from (\ref{eq:sim}), 
where the periodic part of the Bloch functions have not been explicitly 
included, consistent with not explicitly including the atomic orbitals in the 
TB description. Note that the oscillatory behavior coming from the interference 
among the plane-wave part of the six $\phi_\mu$ is well captured by the TB 
envelope function. 

The good agreement between TB and EMT is limited to distances from the 
impurity site larger than a few lattice parameters ($\sim$~1~nm). Closer 
to the impurity, particularly at the impurity site, the TB results become 
much larger than the EMT prediction, in qualitative agreement with experiment. 
\cite{emtsi} This reflects central cell effects, not included in the EMT 
expressions (\ref{eq:sim}) and (\ref{eq:envelope}). In the central cell region, 
the discrepancy between TB and EMT wavefunctions is significantly larger than 
those reported for donors in GaAs, \cite{martins02} a result that could have 
been anticipated from the spectral weight given in Fig.~\ref{fig1}(b). 
EMT rests on the assumption that the impurity eigenstate is highly 
localized in $k$-space, so that only Bloch states near the conduction 
band minima enter in the expansion, as implied in Eq.~(\ref{eq:sim}).
This is the case for GaAs,\cite{martins02} where for a range of values of 
$U_0$ ($U_0<1.8$ eV) we find $W(\Gamma)$ essentially equal to one, in agreement 
with the EMT assumption. In Si, even small values of $U_0$ yield spectral 
weights at $\Delta_{min}$ well below one. For $U_0=U_{\rm P}$ in particular, 
$W(\Delta_{min})\cong 0.3$.

We remark that the sharp shallow-to-deep transition obtained for GaAs in 
Ref.~\onlinecite{martins02}, with kinks in the curves of $E_b$ and $W$ 
versus $U_0$, is not reproduced here (see Fig.~\ref{fig1}). We attribute this 
to the lack of a strictly shallow region, with the spectral weight of the donor 
state concentrated in one or a few $k$-points. Therefore, while the binding 
energy of shallow donors in GaAs is essentially constant, independent of the 
species ($\sim 6$~meV for C, Si and Ge, in excellent agreement with the EMT 
estimate), in Si it varies according to the donor species (45 meV for P, 53 meV 
for As and 42 meV for Sb, to be compared with the EMT estimate of 30 meV). It is 
interesting to note in Fig.~\ref{fig1}(a) that, as  the impurity level 
becomes shallower by decreasing $U_0$, $E_b$ approaches the EMT 
estimate for the binding energy.\cite{emt}  

\section{Donors under an uniform Electric Field}
\label{sec:field}

The formalism presented in Sec.\ref{sec:TB} is easily extended to include 
an uniform electric field in the system. Assuming a constant field $\mathbf{E}$ 
applied along the $[00\bar 1]$ direction, it is incorporated in the TB 
formalism by modifying the on-site energies in Eq.~(\ref{hamilton}) as 
follows:\cite{graf,ribeiro}
\begin{equation}
h_{ii}^{\nu\nu}(E) = h_{ii}^{\nu\nu}(0) - |e|Ez_{i}.
\label{campo}
\end{equation}
Periodic boundary conditions lead to a discontinuity in the 
potential at the supercell boundary $z_i=Z_B$, where $Z_B$ is half of the 
supercell length along [001] or, equivalently, the distance from the 
impurity to the Si/barrier interface. The potential discontinuity, $V_B=2|e|EZ_B$, 
actually has a physical meaning in the present study: It models the potential 
due to the barrier material layer above the Si host\cite{kane} (see inset in Fig.~\ref{fig3}).

A description for the A-gate operations may be inferred 
from the behavior of the TB envelope function squared at the impurity site 
under applied field $E$, normalized to the zero-field value:
\begin{equation}
A/A_0 = |\Psi_{EF}^E(0)|^2/|\Psi_{EF}^0(0)|^2
\label{ratio}
\end{equation}
The notation here indicates that this ratio should follow a behavior similar 
to that for the hyperfine coupling constants between the donor nucleus and 
electron with $(A)$ and without $(A_0)$ external field. Since the 
hyperfine interaction $A$ is proportional to $|\Psi (0)|^2$, and  
we are using here the envelope rather than the full TB eigenfunctions, this  
equivalence is not rigorous. The ratio in (\ref{ratio}) is plotted in 
Fig.~\ref{fig3}(a) for three values of the impurity depth with respect to the 
Si/barrier interface. Calculations for $Z_B$=10.86~nm were performed with 
cubic supercells $(L=40)$, while for $Z_B$= 5.43 and 21.72~nm tetragonal 
supercells with $L_x=L_y=40$ and $L_z=20$ and 80 respectively were used.
At small field values we obtain a quadratic decay of $A/A_0$ with $E$, 
in agreement with the perturbation theory results for the hydrogen atom.
\cite{schiff} At large enough fields, $|\Psi_{EF}^E(0)|^2$ becomes vanishingly 
small, and the transition between the two regimes is qualitatively different
according to $Z_B$: For the largest values of $Z_B$ we get an abrupt transition 
at a critical field $E_c$, while smaller $Z_B$ (e.g. $Z_B=5.43$~nm) lead to a 
smooth decay, similar to the one depicted in Ref.~\onlinecite{kane}. In this latter 
case, we define $E_c$ as the field for which the curve $A/A_0$ vs $E$ has an 
inflection point, where $A/A_0\sim 0.5$, thus $E_c(5.43 \rm{nm}) = 130$ kV/cm.
We find that the decrease of $E_c$ with $Z_B$ follows a simple rule 
$E_c \propto 1/Z_B$, as given by the solid line in Fig.~\ref{fig3}(b).

In order to analyze the different regimes illustrated in 
Fig.~\ref{fig3}(a), we study the overall behavior of the envelope squared 
profile along the $z$-axis,
\begin{equation}
\rho (z) =\sum_{s=1}^2\sum_{x_i^s y_i^s}|\Psi_{EF}^E(\mathbf{R}^{s}_i)|^2,
\label{profile}
\end{equation}
where the first summation is over the two fcc sublattices, with 
$\mathbf{R}^{s}_i$ corresponding to the atomic sites in sublattice $s$, thus
$z$ labels each monolayer in the diamond structure, and $\rho(z)$
quantifies the $z$-projected charge distribution for the electron states under 
applied field $E$. Fig.~\ref{fig4} gives $\rho(z)$ for the electron ground 
state and also for the first excited state with $Z_B=10.86$~nm as the applied 
field increases. Up to fields very close to $E_c~(\sim 53$~kV/cm), the ground 
state distribution retains essentially the bound donor character, with the 
electronic charge accumulating predominantly around the impurity $(z=0)$. For 
$E>E_c$ we observe an abrupt charge transfer towards the barrier, with some residual 
charge remaining at the impurity site. The first excited state displays a 
complementary behavior, with charge transfer from the barrier into the impurity 
region as $E$ increases. The binding energies (energy eigenvalues relative to the 
bottom of the conduction band) are calculated here taking into account 
the dependence of the conduction band edge under applied field. The binding energies 
of the two lowest electron states are given in Fig.~\ref{fig5}(a). Note that they 
cross at $E_c$.
 
The binding energies of the two lowest eigenstates for $Z_B=5.43$~nm are presented in 
Fig.~\ref{fig5}(b). They do not cross, but rather display an anticrossing 
behavior, confirmed by the corresponding doubled-peaked charge distributions in 
Fig.~\ref{fig6}, with wavefunctions extending over the attractive wells 
of  the impurity and of the electric field potential. This is consistent of eigenstates 
which are superpositions of bound states in each potential well. Note that for $E=E_c$ in 
Fig.~\ref{fig6}(c), the two states have essentially the same charge distribution,
as expected at the anticrossing point. The anticrossing in Fig.~\ref{fig5}(b) is such that 
for $E<E_c$ the lines giving the two states are essentially parallel, converging 
asymptotically at zero field  to the binding energies 45.6 meV, for the $A_1$ ground 
state, and 32.4 meV for the first excited state. This is very close to the experimental 
binding energy of the excited $E$ (32.6 meV) and $T_2$ (33.9 meV) states, which can not 
be individually resolved within our variational scheme.\cite{foot-symmetry} Note 
that this was independently obtained with the same value of the parameter $U_0$, 
chosen to fit the $A_1$ state binding energy alone. Near and above $E_c$ a typical 
2-level anticrossing behavior is obtained, with the excited state eventually merging 
into the conduction band at $E=150$ kV/cm.

The above results may be understood within a simple picture of the electron  
in a double well potential, the first well being most attractive at the 
impurity site, $V(\mathbf{R}=0)=-U_0$, and the second well at the barrier 
interface, $V(z=Z_B)= -V_B/2=-|e|EZ_B$ neglecting the Coulomb potential 
contribution (\ref{potential}) at the interface. An internal barrier separates 
the two wells and, for a fixed $E$, this internal barrier height and width 
increase with $Z_B$. Deep donor positioning leads to a weaker coupling between 
the states localized at each well, even close to level degeneracy, resulting the 
level crossing behavior illustrated in Fig.\ref{fig5}(a). For donor positioning 
closer to the interface the internal barrier gets weaker, enhancing the coupling 
between levels localized in each well and leading to wavefunction superposition  
and to the anticrossing behavior illustrated in Fig.\ref{fig5}(b). The scaling of 
$E_c$ with $1/Z_B$ may also be understood assuming that the critical field 
corresponds to the crossing of the ground state energies of two wells: The 
Coulomb well and an approximately triangular well at the barrier. Since the 
relative depths of the wells increases with $EZ_B$, and assuming that the ground 
states energies are fixed with respect to each well's depth, leads to the 
$E_c\propto 1/Z_B$ behavior.

\section{Adiabatic processes driven by an uniform Electric Field}
\label{sec:times}

Coherent manipulation of electrons by the A-gates requires 
that the switching time between different electron states be slow enough to 
guarantee adiabaticity of the process. Instantaneous eigenstates of $H(t)$ may 
thus be defined at any time $t$. In the present case, we assume a linear 
increase of the external field from 0 to a maximum value $E_{\rm max}$ so that 
$H(t)= H(0)-|e|E_{\rm max} z t$, with $0<t<T$, where $T$ is the total switching 
time. A lower bound for $T$ is obtained from the adiabatic theorem,
\cite{messiah,farhi} following Ref.\onlinecite{ribeiro}:
\begin{equation}
T_a=\frac{\hbar|e|E_{\rm max} Z_B}{g_{\rm min}^2},
\label{adiabatic}
\end{equation}  
where $g_{\rm min}$ is the minimum gap between the two lowest electron 
states.  In the anticrossing case illustrated in Fig.\ref{fig5}(b), we 
get $g_{\rm min}\cong 9.8$ meV. Assuming that a totally ionized state is 
required as the final state, we take $E_{\rm max} = 180$ kV/cm, leading to 
$T_a\sim0.5$~ps. This is a perfectly acceptable time for the operation of 
A-gates in spin-based Si QC, given the relatively long electron spin coherence 
times (of the order of a few ms) in Si.\cite{sousa}

As the impurity distance from the barrier increases, one eventually reaches 
the crossing regime, when $g_{\rm min}\to 0$, meaning that $T_a\to\infty$ 
and no adiabatic ionization is possible. Ionization would still occur for 
$E>E_c$, but as a stochastic decay process from the first excited 
state. From Fig.~\ref{fig3}(a) we see that the A-gate might be used to 
partially reduce the contact interaction, in the case of $Z_B=10.86$~nm to 
about 20\% of its value at zero field. For larger $Z_B$ the range for 
adiabatic variation in $A/A_0$ is even smaller. Therefore $Z_B\sim 5$~nm seems 
to be a favorable positioning for the donors, since it allows adiabatic 
reduction of $A/A_0$ to any desired final value, with this ratio varying 
smoothly from one (at $E=0$) to zero (for $E=E_{\rm max}\sim 2E_c$).

\section{Summary and Conclusions}
\label{sec:end}

We have presented a TB study of donor levels in Si. The reliability of 
the TB approach for the present study was verified by comparison of the 
TB and EMT envelope functions as well as by the value predicted for the 
$A_1 - \{E, T_2\}$ energy splitting in agreement with experiment within 
our numerical accuracy. Previous TB studies of intermediate and shallow 
impurity levels in semiconductors\cite{form1,form2,martins02} dealt with 
materials with band extrema at ${\bf k}= 0$, and the present results show 
that the oscillatory behavior of the wave function due to interference 
effects in the plane-wave part of the Bloch wavefunctions, typical of 
degenerate band extrema at $\bf k \ne 0$, is well is well captured 
by the TB approach.

\begin{figure}
\includegraphics[height=6in]{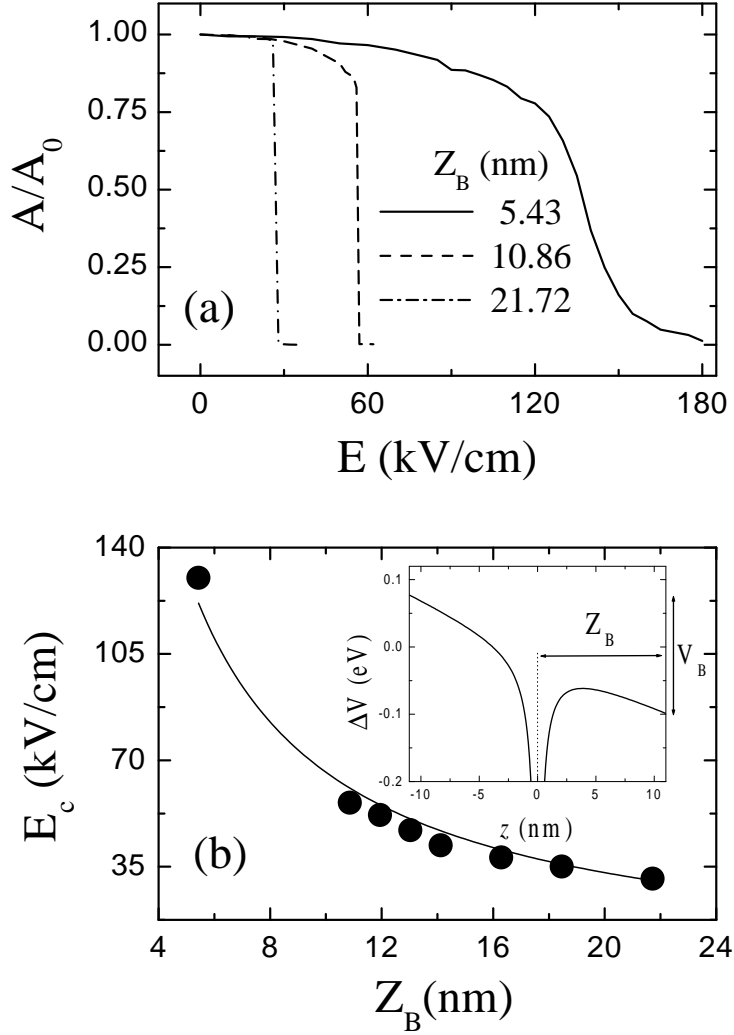}
\caption{\textbf{(a)} TB envelope function squared at the impurity site 
under applied field $E$, normalized to the zero-field value, for the indicated 
values of the impurity-Si/barrier interface distance $Z_B$. \textbf{(b)} Dependence 
of the critical field $E_c$ on $Z_B$. The solid line is a best fit of the form 
$E_c \propto 1/Z_B$. The inset gives a schematic representation of the perturbation 
potential added to the bulk Si hamiltonian due to the impurity at ${\bf R}=0$ and 
to a uniform electric-field in the negative $z$ direction.
\label{fig3}}
\end{figure}

\begin{figure}
\includegraphics[height=6in]{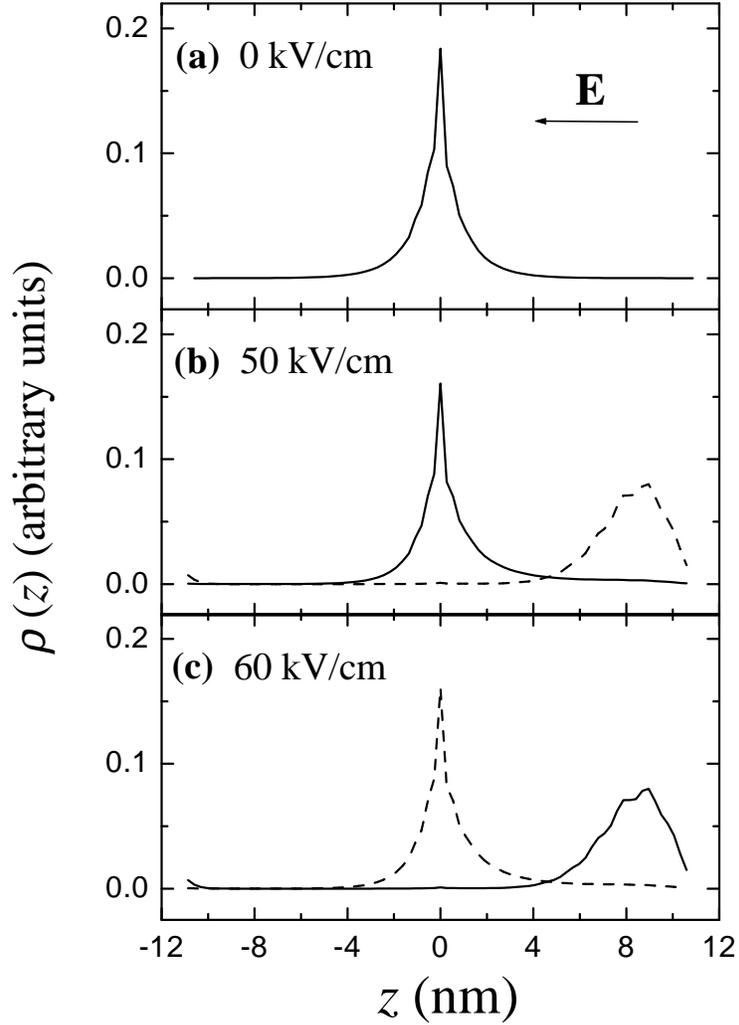}
\caption{Tight binding envelope function squared projected along $z$ for  
$Z_{B}=10.86$~nm and the indicated values of the field $E$ applied in the 
negative $z$ direction. The solid (dashed) line gives the donor ground 
(1st excited) state. Note in (b) and (c) the exchange among the $\rho(z)$ 
for the lowest energy states [{\it ground}] $\leftrightarrow$ [{\it excited}] 
which occurs over a narrow rage of electric field increase, a signature of the 
crossing behavior in Fig.~\ref{fig5}(a).
\label{fig4}}
\end{figure}

\begin{figure}
\includegraphics[height=6in]{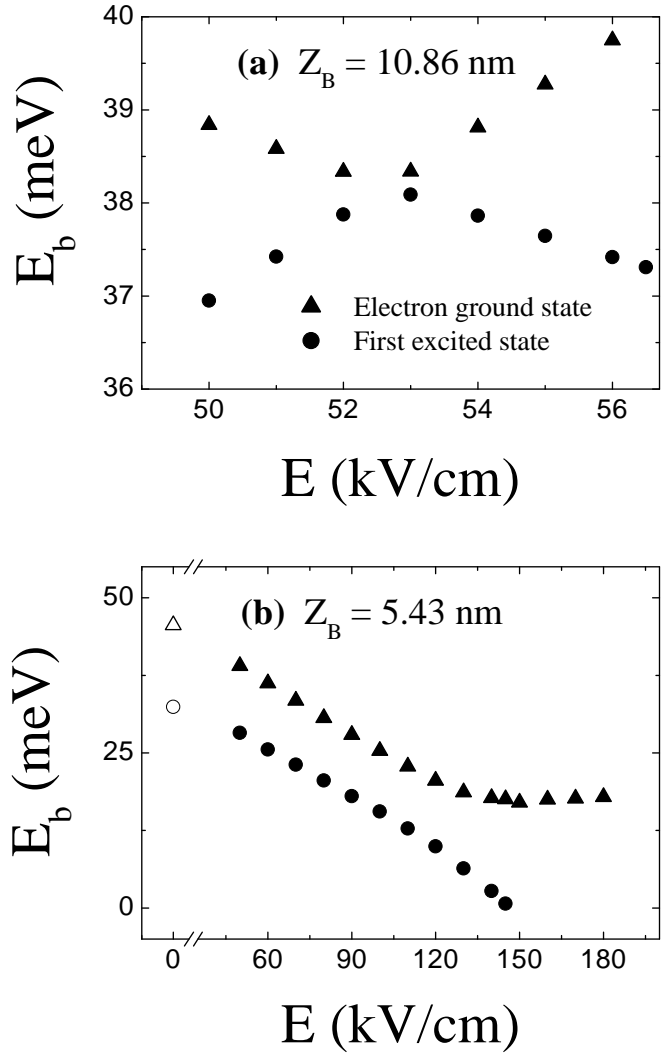}
\caption{Calculated binding energies versus electric field intensity 
of the two lowest donor electron states. \textbf{(a)} For $Z_{B}=10.86$~nm 
the energies reveal a crossing regime. \textbf{(b)} Anticrossing of the 
two lowest electron states for $Z_{B}=5.43$~nm. The open symbols correspond 
the zero field calculated values: 45.6 meV and 32.4 meV.
\label{fig5}}
\end{figure}

\begin{figure}
\includegraphics[height=6in]{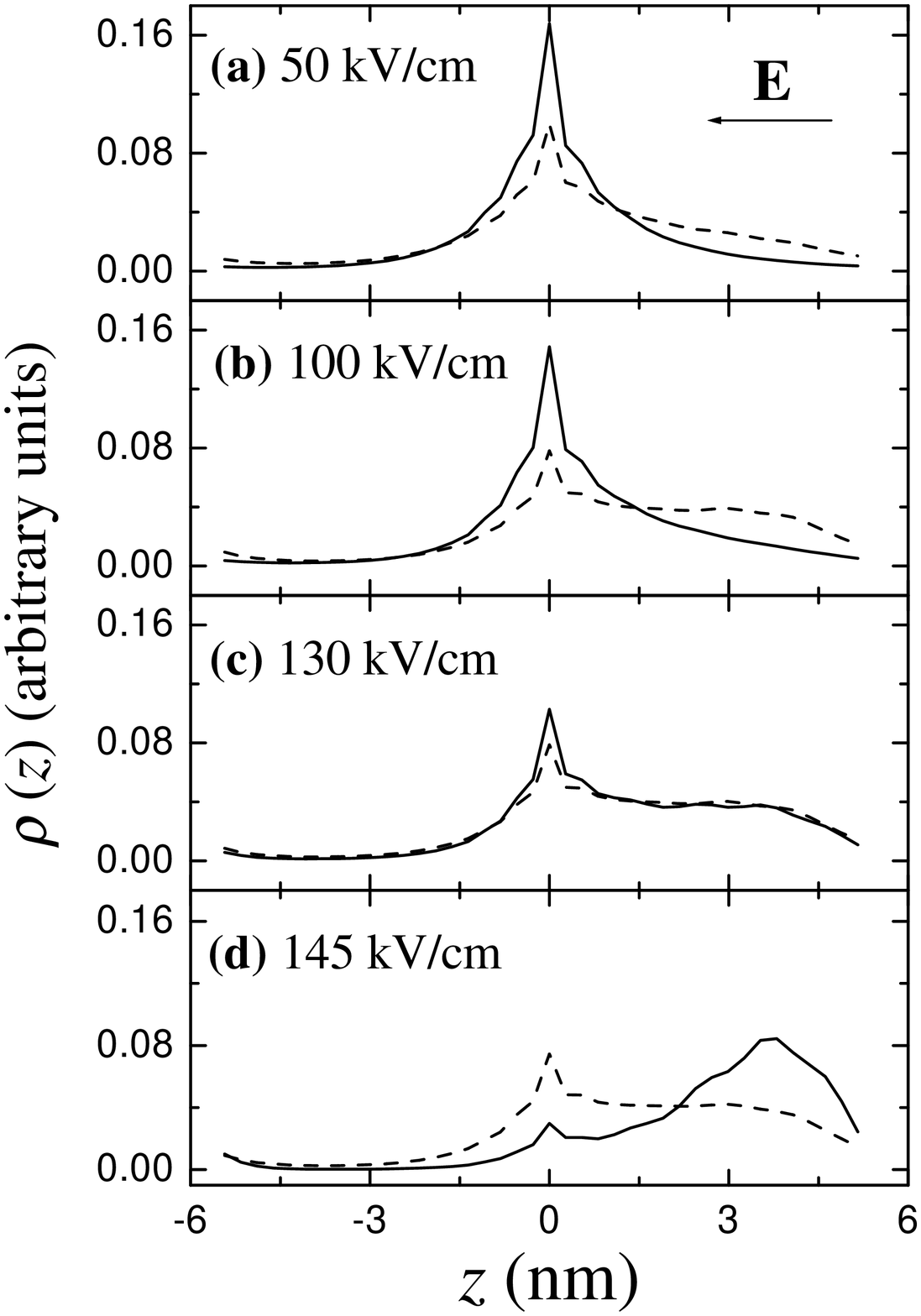}
\caption{TB envelope function squared projected along $z$ for  
$Z_{B}=5.43$~nm and the indicated values of the applied field $E$. 
The solid (dashed) line gives the ground (1st excited) state.
At the critical field in {\bf (c)} the two states have similar 
charge distributions, typical of a superposition of states localized in 
each well and a signature of the anticrossing behavior
in Fig.~\ref{fig5}(b).
\label{fig6}}
\end{figure}

In the presence of an increasing uniform electric field, the donor states 
respond in different ways according to the donor depth $Z_B$ below the 
Si/barrier interface. For deeply positioned donors, i.e, for $Z_B>>a,b$, 
where $a$ and $b$ are the Bohr radii for P in Si, abrupt ionization occurs 
at a critical field $E_c$, while for $Z_B$ greater but of magnitude comparable 
to the Bohr radii, a smooth electronic charge transfer from the donor site
towards the barrier interface is obtained, eventually leading to complete 
ionization. The different regimes were identified in three ways: (i) From the 
decrease in electronic charge at the donor nucleus [Fig.~\ref{fig3}(a)]. This 
behavior implies an analogous dependence of the electron-nucleus hyperfine 
coupling constant $A$ as a function of the increasing external field. (ii) From 
charge distributions (Figs.~\ref{fig4} and \ref{fig6}), where the superposition of 
donor-like and barrier-like bound states is inhibited for deeply positioned donors.  
(iii)  From the behavior of the binding energies of the two lowest electron states 
as the applied field increases (Fig.~\ref{fig5}), changing from a level-crossing into 
an anticrossing regime as $Z_B$ decreases. The donor excited states in the S-like 
manifold also play a role in the anticrossing regime, as illustrated for 
$E\lesssim E_c$  in Fig.~\ref{fig5}(b). 

The minimum gap $g_{\rm min}$ in the anticrossing regime is a key ingredient 
determining the possibility of an adiabatic evolution of the electron state 
under the action of the A-gates. Given that the product $E_c Z_B$ is approximately 
constant [see fit Fig.~\ref{fig3}(b)], the {\it adiabatic} time $T_a$ in 
(\ref{adiabatic}) is expected to depend very weakly in the product $E_{\rm max} Z_B$, 
assuming one aims at complete ionization.\cite{skinner} Therefore $T_a$ should 
not depend explicitly on $Z_B$, but only implicitly through $1/g_{\rm min}^2$. 
We have shown that for $Z_B\sim 5$~nm, i.e., about twice the largest Bohr radius $a$ 
in Eq. (\ref{eq:envelope}), electric field switching times smaller than 1 ps may be reached, 
which is a favorable operation time given the long electronic spin coherence times in Si.
If one aims at a final state where only partial reduction of the electronic 
charge at the nucleus occurs\cite{kane,vrijen}, values of $Z_B$ of this order 
of magnitude are still the most convenient, since any final value of the 
nuclear charge may be attained. 

The Bloch phases interference behavior in the donor wavefunctions has been 
previously shown to lead to oscillatory behavior of the exchange coupling between 
two donors,\cite{koiller01} affecting the two-qubit operations in exchange-based
architectures in Si. We remark that such oscillations are well captured in the 
TB wavefunctions, and that the present study demonstrates that electric field 
control over single donor wavefunctions, such as proposed in A-gate operations, 
\cite{kane,vrijen,skinner} do not present additional complications due 
to the Si band structure. The only critical parameter is the donor positioning 
below the Si/barrier interface, which should be chosen and controlled according 
to the physical criteria presented here.

\acknowledgments
We thank Daniel Loss and Bruce Kane for interesting discussions. 
We also thank Fabio Ribeiro for useful suggestions. This work is partially 
supported by Brazilian agencies Conselho Nacional de Desenvolvimento Cient\'\i fico 
e Tecnol\'ogico (CNPq), CAPES, Funda\c c\~ao de Amparo \`a Pesquisa do Rio de 
Janeiro (FAPERJ), Funda\c c\~ao Universit\'aria Jos\'e Bonif\'acio (FUJB),  
PRONEX-MCT, and Instituto do Mil\^enio de Nanoci\^encias-CNPq.

\begin{thebibliography}{100}
\bibitem{foot-symmetry} The six-fold degenerate S-like donor states in Si split into 
three components due to the tetrahedral symmetry of the host: A non-degenerate 
ground state of $A_1$ symmetry given in Eq.~(\ref{eq:sim}), an excited triplet 
of symmetry $T_2$ followed by an excited doublet of symmetry $E$. The TB treatment
adopted here correctly describes the symmetry and ordering of the states within 
this manifold when full diagonalization of the TB Hamiltonian is performed. 
Within the variational scheme we have not resolved the nearly-degenerate 
$T_2$ and $E$ levels, which are only 1.3 meV apart.
\end{thebibliography}
\end{document}